\documentclass[a4paper,preprint]{elsarticle}
\usepackage{graphicx}
\usepackage{subfigure}
\usepackage{amssymb}
\usepackage{amsmath}

\usepackage{microtype}

\usepackage[colorlinks=true]{hyperref}

\usepackage{color}

\newcommand{\R}{\ensuremath{\mathbb{R}}}
\newcommand{\C}{\mathbb{C}}

\newcommand{\W}{\mathcal{W}}

\newcommand{\bigoh}{\mathcal{O}}

\newcommand{\rank}{\ensuremath{\operatorname{rank}}}

\newcommand{\sgn}{\ensuremath{\operatorname{sgn}}}

\definecolor{textaddcolor}{rgb}{0,1,0}
\definecolor{textdelcolor}{rgb}{1,0,0}

\journal{Journal of Computational and Applied Mathematics}
\begin{document}

\begin{frontmatter}
\title{Applying numerical continuation to the parameter dependence of solutions of the Schr\"odinger equation}

\author[ua]{Jan Broeckhove}\ead{jan.broeckhove@ua.ac.be}
\author[ua]{Przemys\l aw K\l osiewicz}\ead{przemyslaw.klosiewicz@ua.ac.be}
\author[ua]{Wim Vanroose}\ead{wim.vanroose@ua.ac.be}

\address[ua]{Department of Mathematics and Computer Science, Universiteit Antwerpen, Middelheimlaan 1, Antwerpen 2020, Belgium}

%%
%% Abstract:
\begin{abstract}
In molecular reactions at the microscopic level the appearance of resonances has an important influence on the reactivity. It is important to predict when a bound state transitions into a resonance and how these transitions depend on various system parameters such as internuclear distances. The dynamics of such systems are described by the time-independent Schr\"odinger equation and the resonances are modeled by poles of the $S$-matrix.

Using numerical continuation methods and bifurcation theory, techniques which find their roots in the study of dynamical systems, we are able to develop efficient and robust methods to study the transitions of bound states into resonances. By applying Keller's \emph{Pseudo-Arclength continuation}, we can minimize the numerical complexity of our algorithm. As continuation methods generally assume smooth and well-behaving functions and the $S$-matrix is neither, special care has been taken to ensure accurate results.

We have successfully applied our approach in a number of model problems involving the radial Schr\"odinger equation.
\end{abstract}

\end{frontmatter}

%% Matter:
%!TEX root = main.tex
\section{Introduction} 
\label{sec:intro}
Over the last couple of decades several reliable numerical methods have been developed for continuation of solutions and bifurcation analysis for dynamical systems \cite{Keller1977,Allgower1992,Doedel2007,Krauskopf2007}. In this contribution we investigate the application of these methods in the context of quantum physics. In particular, we use numerical continuation to trace the dependence of the energy and width of resonances on the system parameters. This is relevant e.g.\ in low energy electron-molecule scattering where the occurrence and structure of the resonance depend on the internuclear distance in the molecule.

Our system of interest fits the radial Schr\"odinger equation, a subclass of Sturm-Liouville boundary value problems.  For these types of problems, there exist several very accurate methods that find the bound state eigenvalues \cite{Pruess1993,Ledoux2005}. In many physical systems, however, it is also valuable for finding the resonant states that have a complex valued energy. 

We define resonances and bound states as solutions of the Schr\"odinger equation for an energy where the $S$-matrix has a pole \cite{Taylor2006,Newton1982,Burke1995}. The $S$-matrix is a function of the complex momentum $k$ that can be extracted from the solution at the end of the domain. It also depends on the system parameters. We introduce a regularization procedure that transforms the poles into zeros and smoothes the behavior near the origin in the $k$-plane. This allows the application of the pseudo-arclength continuation method to trace the trajectory of the zeros, and hence the poles, as the system parameter changes.

The outline of the paper is as follows. In section \ref{sec:numcont} we present an overview of the concepts underlying the numerical method that constructs a solution set of a non-linear equation with the help of numerical continuation. As indicated in the application in section \ref{sec:results}, we use an implementation of these methods provided by the AUTO package \cite{Keller2003}. In section \ref{sec:qm} we review the concepts related to the Schr\"odinger equation, its solution through the renormalized Numerov method and the extraction of the $S$-matrix from the numerical wave function. It is the poles of the $S$-matrix that are subjected to the numerical continuation methods of section \ref{sec:numcont}.  Finally, in section \ref{sec:results} we demonstrate our approach on two models describing a single-particle in three dimensions in a spherically-symmetric potential. Using partial wave expansion, these scattering problems reduce to a radial Schr\"odinger equation.

%!TEX root = main.tex
\section{Numerical continuation methods}
\label{sec:numcont}
Numerical continuation methods approximate the solution set of some non-linear equation $F(\mathbf{u},\lambda)=\mathbf{0}$ that depends on a system parameter $\lambda$:
\begin{eqnarray}
	F:\R^{n+1} &\longrightarrow \R^{n} : (\mathbf{u},\lambda) &\longmapsto F(\mathbf{u},\lambda),
\end{eqnarray}
where $\mathbf{u}\in\R^{n}$.  The Implicit Function Theorem states that under certain continuity conditions the solution set is a one-dimensional manifold and can be parameterized by some real parameter $s$. The choice of that parameter is an important one and depends on the method used. Generally, we are
interested in the evolution of the solutions $\mathbf{u}$ in terms of $\lambda$ and this suggests to take $\lambda$ as the continuation parameter. However, this choice may result in difficulties when the solution path passes through a fold. The pseudo-arclength continuation~\cite{Keller1977}  deals with these situations gracefully.

We introduce several notations used throughout this paper. When the distinction between the function variables $\mathbf{u}$ and the parameter $\lambda$ is irrelevant, we write $\mathbf{x}=(\mathbf{u},\lambda)$ and $\mathbf{x}_i=(\mathbf{u}_i,\lambda_i)$ for the subsequent points on the solution curve. The continuation curve is denoted by
$\mathbf{x}(s)$, which emphasizes the dependence on the continuation parameter $s$. The initial point on the curve is associated with $s=0$ and written as $\mathbf{x}_0=\mathbf{x}(0)$. Numerical continuation methods use this point on the curve, along with an initial direction of continuation to construct a sequence of points
\begin{equation}
	\left\{ \mathbf{x}_i \ | \ i=0,\ldots,N \ \textrm{ and } \ F(\mathbf{x}_i)=\mathbf{0} \right\},
\end{equation}
that approximates the solution curve.

\subsection{Pseudo-Arclength Continuation}
The algorithm follows a \emph{predictor-corrector} scheme to construct, starting from an initial solution point $\mathbf{x}_0$, the successive points on the solution curve.

The predictor step is an \emph{Euler} predictor that uses the unit length tangent vector $\dot{\mathbf{x}}_i$ to the curve at a solution point $\mathbf{x}_i$ (thus satisfying $F(\mathbf{x}_i)=\mathbf{0}$) and a \emph{step size} $\Delta s$ to predict a guess
$\mathbf{x}^p_{i+1}$ for the next point on the curve:
\begin{equation}
	\label{eq:predictor}
	\mathbf{x}^p_{i+1} = \mathbf{x}_i + \Delta s \dot{\mathbf{x}}_i.
\end{equation}

\begin{figure}
	\centering
	\includegraphics[width=5.0cm]{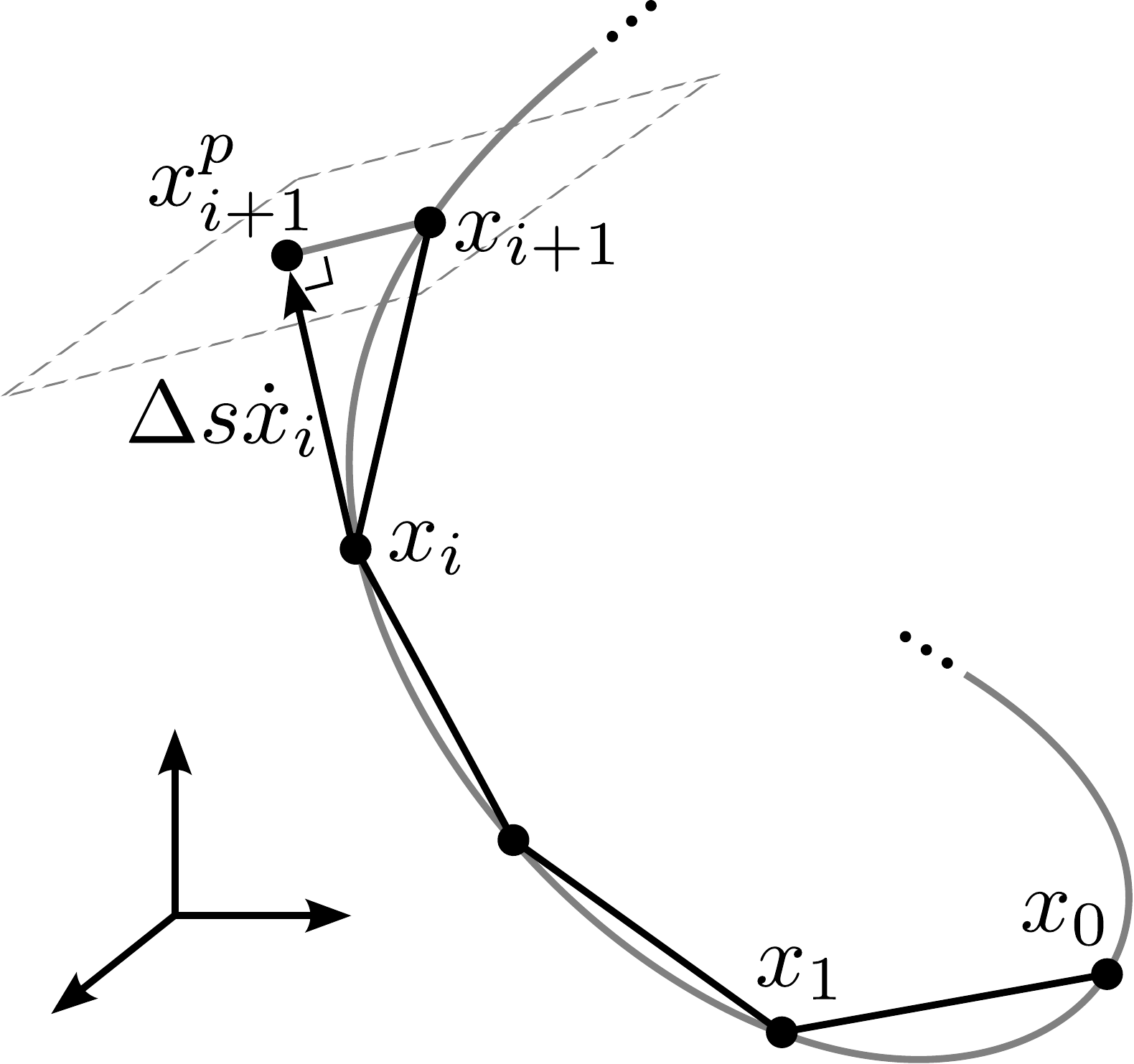}
	\caption{Graphical representation of one predictor-corrector step in pseudo-arclength continuation as discussed in section \ref{sec:numcont}.}
	\label{fig:pac}
\end{figure}

The corrector step improves the guess $\mathbf{x}^p_{i+1}$ with a Newton iteration on the \emph{augmented} system to obtain a new solution point $\mathbf{x}_{i+1}$.  This augmented system has, in addition to the constraint $F(\mathbf{x}_{i+1}) = \mathbf{0}$, the requirement that $\mathbf{x}_{i+1}$ must lie on the hyperplane through $\mathbf{x}^p_{i+1}$ perpendicular to $\dot{\mathbf{x}}_i$, the tangent to the previous solution. This translates to 
\begin{equation}
	\label{eq:corrector}
	\left\{
		\begin{array}{l}
			F(\mathbf{x}_{i+1}) = \mathbf{0} \\
			\left( \mathbf{x}_{i+1}-\mathbf{x}^p_{i+1} \right) \cdot 		\dot{\mathbf{x}}_{i} = 0.
		\end{array}
	\right.
\end{equation}
This system is a map from $\mathbb{R}^{n+1}$ to $\mathbb{R}^{n+1}$ and defines, under some conditions that are usually met, uniquely the next point on the solution curve. It is the point of intersection between the hyperplane and the curve shown in figure \ref{fig:pac}. These steps are common to other Euler-Newton like methods and other approaches to define the next point on the curve are discussed in \cite{Deuflhard2004}.

The tangent vector for the next step is computed by solving:
\begin{equation}
	\label{eq:tangent_system}
	\left(\begin{array}{cc}
		F_\mathbf{u}(\mathbf{x}_{i+1}) & F_{\lambda}(\mathbf{x}_{i+1}) \\
		\dot{\mathbf{u}}^T_i & \dot{\lambda}_i
	\end{array}\right)
	\left(\begin{array}{c}
		\dot{\mathbf{u}}_{i+1} \\
		\dot{\lambda}_{i+1}
	\end{array}\right)
	=\left(\begin{array}{c}
		\mathbf{0} \\
		1
	\end{array}\right),
\end{equation}
and normalizing $||\dot{\mathbf{x}}_{i+1}||=1$. The right-hand side of equation (\ref{eq:tangent_system}) is a column vector consisting of zeros except on the last row.

Note that the Jacobian, $F_{\mathbf{x}}(\mathbf{x}_{i+1})$, is required both for the calculation of the tangent direction $\dot{\mathbf{x}}_{i+1}$ and for the calculation of the Newton corrections.  In our application we only have $F$ numerically so we need to approximate the Jacobian. This is done using finite differences. The $j$th column of the Jacobian matrix is found by a central difference and requires two solutions with slightly different arguments:
\begin{equation}
	(F_\mathbf{x}(\mathbf{x}_{i+1}))_j = \frac{F(\mathbf{x}_{i+1} + \epsilon \mathbf{e}_j) - F(\mathbf{x}_{i+1} - \epsilon \mathbf{e}_j)}{2\epsilon},
\end{equation}
where is $\mathbf{e}_j$ is the $j$th unit vector. A discussion on the optimal choice of $\epsilon$ given the machine precision is found in \cite{Kelley1995}.

\subsection{Regular and singular solutions}
An important notion is the regularity of a solution point \cite{Krauskopf2007,Doedel2007}: a point $\mathbf{x}_i\in\R^{n+1}$ on the solution curve is a \emph{regular} solution of $F(\mathbf{x})=\mathbf{0}$ if the Jacobian matrix, $F_\mathbf{x}(\mathbf{x}_i)$, has maximal rank. Otherwise the solution is singular. Since $F_\mathbf{x}(\mathbf{x}_i)$ has $n$ rows and $n+1$ columns, its maximal rank is $n$.

Another important notion is bifurcation. The solution is said to \emph{bifurcate} \cite{Keller1969} from the solution $\mathbf{u}_t$ at a parameter value $\lambda_t$ if there are two or more distinct solutions which approach $\mathbf{u}_t$ as $\lambda$ tends to a threshold value $\lambda_t$. A more rigorous definition of a bifurcation point can be found in \cite{Allgower1990}.

The connection between these two definitions is that a bifurcation point $\mathbf{x}_i$ of $F(\mathbf{x})=\mathbf{0}$ must be a singular solution which means that:
\begin{equation}
	\rank(F_\mathbf{x}(\mathbf{x}_i)) < n,
\end{equation}
and consequently (rank-nullity theorem):
\begin{equation}
\label{eq:dimkergeq2}
	\dim\ker(F_\mathbf{x}(\mathbf{x}_i)) \geq 2.
\end{equation}
In case the equality in (\ref{eq:dimkergeq2}) holds, we call $\mathbf{x}_i$ a \emph{simple bifurcation point}~\cite{Allgower1990}. We assume this is the only type of bifurcation that occurs in the systems we study here.

Following \cite{Allgower1990} we detect these bifurcation points by looking at the sign of the determinant of the \emph{augmented} Jacobian matrix. When traversing a solution branch a simple bifurcation point lies between two solutions $\mathbf{x}_i$ and $\mathbf{x}_{i+1}$ if and only if
\begin{equation}
\label{eq:diffdetsign}
	\sgn\det\left(\begin{array}{c} F_{\mathbf{x}}(\mathbf{x}_i) \\ \dot{\mathbf{x}}^{T}_i \end{array}\right)
	\quad\neq\quad
	\sgn\det\left(\begin{array}{c} F_{\mathbf{x}}(\mathbf{x}_{i+1}) \\ \dot{\mathbf{x}}^{T}_{i+1} \end{array}\right).
\end{equation}
This allows to find the bifurcation point accurately with a straightforward yet rather slow convergence procedure. Note that $\mathbf{x}_i$ and $\mathbf{x}_{i+1}$ must be close to each other to avoid ``overshooting'' bifurcation points.

\subsection{Branching}
When two solution curves meet in a simple bifurcation point $\mathbf{x}_t$, the dimension of the nullspace of
$F_\mathbf{x}(\mathbf{x}_t)$ is two.  This nullspace is then spanned by two orthonormal vectors $t_1$ and $t_2$.  At the same time, the left nullspace of $F_\mathbf{x}(\mathbf{x}_t)$ is one dimensional since $F$ is a function from $\mathbb{R}^{n+1}$ to $\mathbb{R}^n$. It is spanned by a vector $n_1$.

The two tangent vectors to the curves that depart from the bifurcation point can now be written as a linear combination $\dot{\mathbf{x}}_t = \alpha t_1 + \beta t_2$ of the vectors that span the nullspace. Since the curves fit $F(\mathbf{x}(s))=0$, we can differentiate twice to $s$ and find that the tangent directions fit
\begin{equation}
	F_{\mathbf{xx}} \dot{\mathbf{x}}_t\dot{\mathbf{x}}_t + F_{\mathbf{x}} \ddot{\mathbf{x}}_t = \mathbf{0}. 
\end{equation}
Projection on $n_1$ leads to the {\it algebraic bifurcation equation} \cite{Doedel2007,Mei2000}: 
\begin{equation}
	C_{11} \alpha^2 + 2C_{12} \alpha \beta + C_{22} \beta^2 = 0, \label{eq:abe}
\end{equation}
with $C_{11} = n_1^T F_{\mathbf{xx}} t_1 t_1$, $C_{12} = n_1^T
F_{\mathbf{xx}} t_1 t_2$ and $C_{22} = n_1^T F_{\mathbf{xx}} t_2 t_2$.
In addition we have $\alpha^2 + \beta^2=1$, since the tangent vectors are normalized.

The construction of this equation requires a numerical calculation of the Hessian in the bifurcation point and the determination of the vectors that span the nullspaces.  The solution of the algebraic bifurcation equation gives us the tangent vectors to the curves that depart from the bifurcation point.

\subsection{Implementation}
During initial prototyping we have implemented the above methods in \textsc{Matlab}. For the development of production code we have relied on the well-known implementation of these algorithms in the AUTO package~\cite{Keller2003,AUTO2007}. An alternative implementation is available in the LOCA package which is part of the Trilinos project~\cite{Heroux2005}.

%!TEX root = main.tex
\section{Quantum scattering concepts}
\label{sec:qm}

In this section we review some of the concepts related to the solution of the time-independent Schr\"odinger equation through partial wave analysis and to the $S$-matrix and its properties.

\subsection{The radial Schr\"odinger equation and the $S$-matrix}
The time-independent Schr\"odinger equation
\begin{equation}
\label{eq:tise-3d}
		\left( -\frac{1}{2}\triangle + V(\overline{r},\lambda) \right)\psi(\overline{r}) = E\psi(\overline{r}),
\end{equation}
describes the states of a quantum system with potential $V$ at energy $E$. We let the potential depend on a parameter $\lambda$. How the potential depends on the parameter $\lambda$ is arbitrary.  Any choice is acceptable provided the $\lambda$-dependence is smooth. One possible choice is to scale the potential with a strength $\lambda$ as in $V(\overline{r},\lambda) = \lambda V(\overline{r})$.

In almost all physically relevant situations, $V$ is spherically symmetric, i.e. a function of the radial coordinate $r = \left | \overline{r} \right |$ only. One then transforms equation \eqref{eq:tise-3d} to spherical coordinates $(r, \theta, \varphi )$ and applies the method of separation of variables --- partial wave analysis in physics parlance --- to solve as \cite{Courant1966,Arfken2005}:
\begin{equation}
	\psi(\overline{r}) = \sum c_{lm} \frac{\psi_l(r)}{r} Y_{lm}(\theta,\varphi),
\end{equation}
The spherical harmonics $Y_{lm}$ are the solutions to the angular equation that is independent of $V$. The integer $l$ is the angular momentum. For each $l$ the $\psi_l$ is determined by a radial equation of the following form:
\begin{equation}
\label{eq:tise}
		\left( -\frac{1}{2}\frac{d^2}{dr^2} + V(r,\lambda) + \frac{l(l+1)}{2r^2}\right)\psi_l(r) = E\psi_l(r),
\end{equation}
One refers to the sum of $V$ and the $l$-dependent term as the effective potential.  This equation belongs to a subclass of Sturm-Liouville boundary value problems with $p(x)=1$, $w(x)=1$ and $q(x)$ equal to the  effective potential.

We assume that for $V(r,\lambda)$ the following holds
\begin{equation}
	\begin{array}{llll}
		V(r,\lambda) &= \bigoh(r^{-3-\epsilon}) &\quad\textrm{ for } r\to\infty &\quad\textrm{and } \epsilon > 0 \\
		V(r,\lambda) &= \bigoh(r^{-2+\epsilon}) &\quad\textrm{ for } r\to 0 &\quad\textrm{and } \epsilon > 0,
	\end{array}
\end{equation}
$V$ decays faster than $r^{-3}$ at infinity and is less singular than $r^{-2}$ at the origin. The requirement at infinity limits us to so-called short-range potentials. Extending our approach to the class of long-range interactions requires substantial modifications and is an important direction for future work.

The solution $\psi_l$ of (\ref{eq:tise}) needs to fit the homogeneous Dirichlet boundary condition at $r=0$. Because of the short range of the potential, the solution becomes at $r \rightarrow \infty$ a linear combination of the two fundamental solutions of the free Schr\"odinger equation (i.e.\ without potential term $V$)
\begin{equation}
\label{eq:tise_asymk}
	\left( -\frac{d^2}{dr^2} + \frac{l(l+1)}{r^2} - k^2\right)\psi_l(r) = 0,
\end{equation}
where $k=\sqrt{2E} \in\C$ is the complex momentum. The fundamental solutions are the spherical Riccati-Bessel and Riccati-Neumann functions \cite{Messiah1961,Taylor2006}. Thus, in the asymptotic region we have for the solution of (\ref{eq:tise})
\begin{equation}
\label{eq:psi_is_AhBh}
	\psi_l(r) \rightarrow A_l(k,\lambda) \hat{h}^{+}_{l}(kr) + B_l(k,\lambda) \hat{h}^{-}_{l}(kr) \quad\text{for}\quad r \rightarrow \infty,
\end{equation}
with $A_l(k,\lambda),B_l(k,\lambda) \in\C$. These constants depend on the momentum $k$ and system parameter $\lambda$. The $\hat{h}^{\pm}_{l}$ are Riccati-Hankel functions of the first ($+$) and second ($-$) kinds.  These functions behave asymptotically (up to a phase)  as  $e^{ikr}$, an outgoing, and $e^{-ikr}$, an incoming wave.  The solution is then interpreted as a superposition of an incoming ($\hat{h}^{-}_{l}$) and an outgoing ($\hat{h}^{+}_{l}$) wave.

As the solution is only defined up to an overall normalization, we can renormalize it as follows \cite{Taylor2006}:
\begin{equation}
\label{eq:psi_is_Shh}
	\psi_l(r) = \frac{i}{2}\left( \hat{h}^{-}_{l}(kr) - S_{l}(k,\lambda) \hat{h}^{+}_{l}(kr) \right), \,\,\,\mbox{with}\,\,\, S_{l}(k,\lambda) = -\frac{A_l(k,\lambda)}{B_l(k,\lambda)}.
\end{equation}
This introduces the $S$-matrix, a function of the momentum $k$ and depending on $\lambda$. It determines the phase of the outgoing, scattered wave w.r.t.\ the incoming wave.

\subsection{Resonances and bound states as poles of the $S$-matrix}
\label{subsec:spoles}
It is well established and discussed in several textbooks that poles of $S_l(k,\lambda)$ with $\Im(k) > 0$ correspond to bound states with energy $k^2/2$ and poles with $\Im(k)<0$ correspond to resonances. 

Indeed, if $E$ is a negative real number where $S_l$ diverges, it means that the solution is asymptotically a multiple of $\hat{h}^+_l$ only. Since $E<0$, the momentum $k$ is purely imaginary and $\hat{h}^+_l$ becomes a decaying exponential that  fits the zero boundary conditions as $r \rightarrow \infty$. The solution then fits the boundary conditions $\psi(0)=0$ and $\psi(\infty)=0$ and is then a bound state solution of the boundary value problem.

On the other hand, if $E$ is a complex number with $\Im(k)<0$ where $S_l$ has a pole, the state is classified as a resonance.  Again the asymptotic solution is a multiple of $\hat{h}^+_l$, an outgoing oscillating wave, only.

 Furthermore, as the system parameters change, the poles can move from the upper part towards the lower of the complex plane along a continuous curve and the solutions then transition from bound to resonant state.  A thorough discussion on such trajectories of poles is given in \cite{Taylor2006,Newton1982,Burke1995}.
\begin{figure}
	\centering
	\includegraphics[width=6cm]{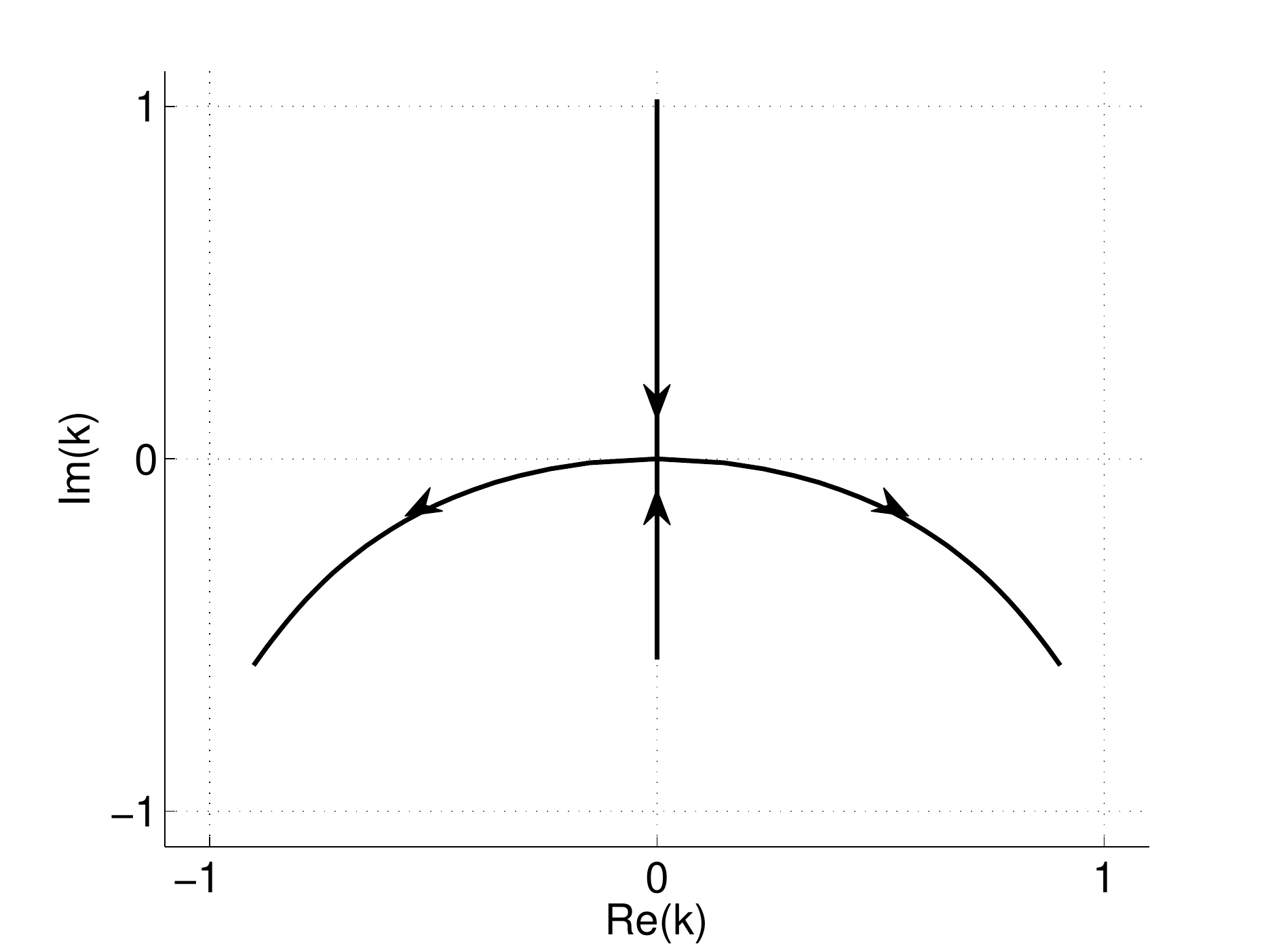}
	\caption{Schematic trajectories of the poles of the $S$-matrix for the Gaussian well example with $l=1$. The four parts of the solution meet in the origin of the complex $k$-plane for $\lambda\approx6.0497$. In this article this point is interpreted as a bifurcation point.}
        \label{fig:pole-trajectory}
\end{figure}

In mathematical terms, we are faced with the problem of finding $k(\lambda):\mathbb{R} \rightarrow \mathbb{C}$, the
solution set of
\begin{equation}
	S_l(k,\lambda)^{-1}=0.
	\label{eq:poles_of_s_matrix}
\end{equation}
When we treat its real and imaginary parts as independent variables, $S_l(k,\lambda)$ is a function from $\mathbb{R}^3$ to
$\mathbb{R}^2$.  It is a non-linear function and only for a few potentials it is available as an analytical expression. In general, the value of the $S$-matrix for a given $\Re(k)$, $\Im(k)$ and $\lambda$ is only found through the numerical solution of the Schr\"odinger equation with methods  such as the $R$-matrix~\cite{Burke1993}, $J$-matrix~\cite{Alhaidari2008}, ECS~\cite{Rescigno2000}, shooting methods~\cite{Johnson1977} and others. Given a numerical expression for the wave function, we extract the $A_l(k,\lambda)$ and $B_l(k,\lambda)$ from the solution with the help of the Wronskian $\W$ at the asymptotic boundary $R$ of the domain. More specifically, from (\ref{eq:psi_is_AhBh}) we derive:
\begin{align}
	A_{l}(k,\lambda) &= \frac{\W\left(\psi_{l}(R),\,\hat{h}_{l}^{-}(kR)\right)}{\W\left(\hat{h}_{l}^{+}(kR),\,\hat{h}_{l}^{-}(kR)\right)} \\
	B_{l}(k,\lambda) &= -\,\frac{\W\left(\psi_{l}(R),\,\hat{h}_{l}^{+}(kR)\right)}{\W\left(\hat{h}_{l}^{+}(kR),\,\hat{h}_{l}^{-}(kR)\right)}.
\end{align}
This then leads to the following expression for the $S$-matrix:
\begin{equation}
\label{eq:S_and_wronskians}
	S_{l}(k,\lambda) = \frac{\W\left(\psi_{l}(R),\,\hat{h}_{l}^{-}(kR)\right)}{\W\left(\psi_{l}(R),\,\hat{h}_{l}^{+}(kR)\right)},
\end{equation}
that can be computed numerically, provided the first derivative of the wave function at $R$, the asymptotic boundary of the domain, can be computed.

Note that the different solution curves of (\ref{eq:poles_of_s_matrix}) can meet each other in a single point.  For a radial equation with $l=1$, for example, the situation is well understood and we illustrate this with the help of figure \ref{fig:pole-trajectory}. As $\lambda \rightarrow \lambda_t$, the critical system parameter where a bound state becomes a resonance, the  pole moves down on the positive imaginary axis towards the origin.  At the same time, another pole corresponding to the virtual state, approaches the origin along the negative imaginary axis from below.  At $\lambda_t$ the two poles of the $S$-matrix coalesce into a single double pole in  the origin. For system parameters beyond $\lambda_t$ there are again two separate poles corresponding to resonances.  They lie in the third and fourth quadrant of the complex $k$-plane.

\subsection{Transforming poles into zeros}
The numerical method discussed in section~\ref{sec:numcont} requires some smoothness conditions on the function $F(\mathbf{x})$ to work in a reliable and fast way.  To achieve a quadratic convergence rate during the Newton correction it is well known that the Jacobian $F_\mathbf{x}$ needs to be locally Lipschitz continuous.

We intend to apply the continuation method to track the path of the zeros of $S_l(k,\lambda)^{-1}$ as $\lambda$ varies. Unfortunately, this function is meromorphic for the potentials  of interest  \cite{Taylor2006} and does not fit these smoothness requirements, especially for $|k| \ll 1$. Indeed, the scattering matrix has the property that $S_l(k^*,\lambda) = S_l(k,\lambda)^{-1}$, where $k^*$ is the complex conjugate of $k$. This means that if $S_l(k,\lambda)^{-1}$ has a zero in some $k_0$, it will also have a pole in $k_0^*$. And as $\lambda$ approaches $\lambda_t$, $k_0$ moves towards the origin.  In this situation a zero in $k_0$ and a pole in $k_0^*$ approach each other and at the critical parameter $\lambda_t$ they will coalesce.  It is clear that in a
neighbourhood around the critical point, $|k_0| \ll 1$ and $|\lambda-\lambda_t| \ll 1$, the derivatives of $S_l(k,\lambda)^{-1}$ cannot satisfy these smoothness conditions.

In order to desingularize $S_l(k,\lambda)^{-1}$, i.e.\ to avoid this deteriorating behavior as $|k| \rightarrow 0$, we transform to a new function, related to the $S$-matrix, but having polynomial behavior for $|k|\rightarrow 0$. This function is defined as
\begin{equation}
        F_l(k,\lambda) =  \frac{k^{2l+1}}{S_l(k,\lambda)-1}, \label{eq:invregs}
\end{equation}
with $l$ the angular momentum.  It is clear that for $k\neq 0$,  ${F}_l$ will have a zero if and only if $S_l$ has a pole.

Furthermore, we can show that this function is proportional to the Jost function $\mathcal{F}_l(k,\lambda)$, familiar from scattering theory \cite{Taylor2006,Sitenko1991,Amrein1977}. It is related to the ratio of the regular solution $\varphi_l(k,r)$ and the normalized solution $\psi_l(k,r)$ and which is an analytic function for a wide class of potentials and behaves as a polynomial around the origin of the complex plane. To show this proportionality we use equations (11.19) and (12.145) from \cite{Newton1982}. We have that
\begin{equation}
        S_{l}(k,\lambda)=1-4ik^{-1}\int_{0}^{\infty}dr \, kr \, j_{l}(kr)V(r,\lambda)\psi_{l}(k,r),
\end{equation}
where $\psi_l(k,r)$ is the solution to equation (\ref{eq:tise}).  This wave function is proportional to the regular solution $\varphi_l(k,r)$
\begin{equation}
        \psi_l(k,r) =  \frac{k^{l+1} \varphi_l(k,r)}{\mathcal{F}_l(k,\lambda)(2l+1)!!}.
\end{equation}
where $!!$ indicates the double factorial~\cite{Arfken2005}.

For $l=0$ this $\varphi_l(k,r)$ fits equation \eqref{eq:tise} with boundary $\varphi(k,0)=0$ and $\varphi^\prime(k,0)=1$. In the case $l>0$ the condition is: $\lim_{r \to 0}r^{-l-1}\varphi_{l}(k,r)=1$.

With the help of \cite{Newton1982}, it is clear that 
\begin{equation}
        F_l(k,\lambda) =  \frac{k^{2l+1}}{S_l(k,\lambda)-1} =   \frac{\mathcal{F}_l(k,\lambda)}{C},
\end{equation}
where  
\begin{equation}
        C =  \frac{-4i}{k^{l+1}(2l+1)!!}\int_0^\infty\!\!\!\!dr\, kr \,j_l(kr) V(r,\lambda) \varphi_l(k,r) \label{eq:def_C}.
\end{equation}
This constant is bounded.  Indeed, we have the bound from \cite{Taylor2006} and \cite{Newton1955}
\begin{equation}
	| kr j_l(kr)|  \le  C_1 \left(\frac{|k|r}{1+|k|r} \right)^{l+1} e^{|\Im(k)|r},
\end{equation}
and in a similar way from \cite{Newton1982} we have
\begin{equation}
|\varphi_l(k,r)| \le  C_2 \left(\frac{|k|r}{1+|k|r}\right)^{l+1} e^{|\Im(k)| r} |k|^{-l-1},
\end{equation}
with constants $C_1$ and $C_2$.  So we get
\begin{equation}
|C| \le  C_3 \int_0^\infty\!\! dr \left( \frac{r}{1 + |k|r} \right)^{2l+2} |V(r,\lambda)| e^{2|\Im(k)|r}.
\end{equation}
This bound is finite if the integral over the potential is finite. As we can see from (\ref{eq:def_C}) it is clear that as $k \rightarrow 0$, $C$ is only zero for very specific potentials. We conclude that $F_l(k,r)$ is bounded for a wide range of problems.

Note that tracking the zeros of $A_l(k,\lambda)$ or $B_l(k,\lambda)$ is not an alternative since these functions also suffer from the presence of poles.  These poles are removed by taking the $S$-matrix, the ratio of the $A_l$ and $B_l$.

\subsection{The tangent directions in the bifurcation point}
An advantage of working in the $k$-plane, instead of the $E$-plane, is
that the tangents to the solutions that emerge from the bifurcation
point are orthogonal for problems with $l \ge 1$.

Indeed, around $k=0$, the Jost function can be expanded in the form \cite{Newton1982}
\begin{equation}
\mathcal{F}_l(k,\lambda) = a_1(\lambda) + a_2(\lambda)k^2 + \ldots + b_1(\lambda) k^{2l+1} + b_2(\lambda) k^{2l+3} + \ldots  
\end{equation}
where the coefficients are real functions of the system parameter.
Around $(0,0,\lambda_t)$, with $l \ge 1$ this can further written as 
\begin{equation}
	\mathcal{F}(k,\lambda) = \alpha (\lambda-\lambda_t) + \beta k^2 + \mathcal{O}(k^3) + \mathcal{O}\left((\lambda-\lambda_t)^2\right), 
\end{equation}
where $\alpha,\beta \in \mathbb{R}$. When we write $k = x + i y$ and define the function
\begin{equation}
\begin{array}{cccl}
G: &\mathbb{R}^3  &\rightarrow& \mathbb{R}^2\\
   & (x,y,\lambda)	& \mapsto& \left(\Re(\mathcal{F}(x+iy, \lambda)),\Im(\mathcal{F}(x+iy, \lambda))\right),
\end{array}
\end{equation}
this equation $G(x,y,\lambda)  =  0$ follows  the full problem up to order $k^3$ and $(\lambda-\lambda_t)^2$.

The Jacobian in the point $(x,y,\lambda)$ is then 
\begin{equation}
G_\mathbf{x}(\mathbf{x}) = \left( \begin{array}{ccc}
	 2\beta x & -2 \beta y  & \alpha \\	
	2 \beta y  & 2 \beta x & 0 
	     \end{array}
\right),
\end{equation}
which obviously reduces to the following rank one matrix at the bifurcation point $\mathbf{x}_t = (0,0,\lambda_t)$
\begin{equation}
 G_\mathbf{x} = \left( 
\begin{array}{ccc} 
	0  & 0 & \alpha \\
	0  & 0 & 0
\end{array}
\right).
\end{equation}
A basis for the $\mbox{ker}\left(G_\mathbf{x}(\mathbf{x}_t)\right)$ is then 
\begin{equation}
\mbox{ker}(G_\mathbf{x}(\mathbf{x}_t)) = \mbox{span}\left\{\left(\begin{array}{c} 1\\ 0 \\ 0 \end{array}\right), \left( \begin{array}{c} 0 \\1\\ 0\end{array} \right) \right\},
\end{equation}
and for $\mbox{ker}(G_\mathbf{x}^T)$ it is 
\begin{equation}
\mbox{ker}\left({G_\mathbf{x}(\mathbf{x}_t)}^T\right) = \mbox{span}\left\{\left( \begin{array}{c} 0\\ 1\end{array}\right) \right\}.
\end{equation}
The Hessians are then 
\begin{equation}
G_{\mathbf{xx}} = \left(\begin{array}{ccc}
\left(\begin{array}{ccc}
2 & 0 & 0 \\
0 & 2 & 0	
\end{array} \right), & 
\left(\begin{array}{ccc}
0 & -2 & 0 \\
2 & 0 & 0	
\end{array} \right), & 
\left(\begin{array}{ccc}
0 & 0 & 0 \\
0 & 0 & 0	
     \end{array} \right) 
\end{array} \right)
\end{equation}
The coefficients of the algebraic bifurcation equation  \eqref{eq:abe}  are then $C_{11} =  0$, $C_{12} = 2$, $C_{22}  = 0$ what leads to the equation to be solved:
\begin{equation}
	\alpha \beta = 0 \hspace{0.5cm}\mbox{and} \hspace{0.5cm} \alpha^2 + \beta^2 = 1.
\end{equation}
The solutions, which are $(\alpha,\beta) = (0,1)$ or $(1,0)$ then lead to two tangent directions at the bifurcation point: $\dot{\mathbf{x}}_{t,1} = (1,0,0)^T$ and $\dot{\mathbf{x}}_{t,2} = (0,1,0)^T$.
The direction $\dot{\mathbf{x}}_{t,1}$ corresponds to the two resonances that leave along the real axis and $\dot{\mathbf{x}}_{t,2}$ is the direction along the imaginary axis  from which the bound and anti-bound states approach the threshold.   These tangent vectors are indeed orthogonal.

If we would use numerical continuation in the $E$-plane, these two tangent directions would coincide.

Note that AUTO solves, when it detects a bifurcation point, the algebraic bifurcation equation numerically.
%!TEX root = main.tex
\section{Numerical application}
\label{sec:results}

\subsection{Implementation}
For testing purposes we have developed an implementation of the algorithm described. It consists of two main parts:
\begin{enumerate}
	\item A solver for the Schr\"odinger equation and the associated routines to obtain a numerical approximation of the $S$-matrix.
	As indicated in section~\ref{subsec:spoles} many solvers are possible, each suitable for a range of potentials, domains or dimensionality of the problem. For the two examples we present, dealing with the radial Schr\"odinger equation, the Numerov method has proven very successful.
	
	The Numerov method~\cite{Blatt1967,Johnson1977} is a shooting method that exploits the absence of first order terms in the Schr\"odinger equation to arrive at a fairly straightforward algorithm, with equidistant steps, that is of fourth order.
	
	In \cite{Johnson1977} the renormalized Numerov method was proposed, a reformulation of the algorithm in terms of the ratio of the wave function in successive grid points. We have implemented this algorithm in C++ code.
	
	The first derivative of the wave function, required to compute the ratio of the Wronskians in (\ref{eq:S_and_wronskians}) is determined with a formula given in \cite{Johnson1977} and retains the order $\bigoh(h^4)$. We have tested this convergence behavior in our implementation and found that it holds except for potentials with discontinuities, such as the square well. However, this does not prevent the application of the method: it simply lowers the convergence rate.
	\item A routine that performs the numerical continuation process with detection of branches. For this purpose, we use the well-known library for numerical continuation \textsc{AUTO}~\cite{AUTO2007}. The numerical routines for the necessary computation of the Jacobian matrix are also provided by AUTO and use a second order central difference scheme. A comparative study with other continuation libraries is under consideration.
\end{enumerate}

\subsection{Gaussian potential}
As a first model problem we take the third partial wave ($l=3$) in a Gaussian potential well:
\begin{equation}
	V(r,\lambda) = -\lambda e^{-r^2}.
\end{equation}

For potential strength $\lambda=188$ the system has 6 bound states. Decreasing the potential strength pushes these bound state energies towards zero and transforms them successively into resonances. For the renormalized Numerov solver an integration grid $r\in[0,\ 4.8]$ with 8192 points was used. At the end of this interval, the influence of the Gaussian potential is smaller than $10^{-7}$ and is considered negligible. The shooting method is started with the boundary condition at $r=0$, $\psi_l(0)=0$.

The starting points $\mathbf{x}=(k_\Re,k_\Im,\lambda)$ for the six branches were chosen on the positive imaginary axis of the $k$-plane, in a region close to the origin to ensure convergence of the solver. They are presented in table \ref{tab:gaussstartpoints}. To confirm our renormalized Numerov values we have also computed them using the CPM\{16,14\} method implemented in \textsc{matslise}~\cite{Ledoux2005}. There are no significant differences.

Continuation was started in these points in the direction of the origin with an initial prediction step $\Delta s=10^{-2}$ which may vary dynamically between $10^{-4}$ and $5\times10^{-2}$. The critical transition points, where the continuation branches off, were found at the origin of the $k$-plane for threshold values for $\lambda$ given in table \ref{tab:gaussbranchpoints}.

\begin{table}
	\centering
	\footnotesize
	\begin{tabular}{rrrr}
		\hline
		$n$ & $\lambda$ & $\Im(k) (Ren. Numerov)$ & $\Im(k)$ (\textsc{matslise}) \\\hline
		\texttt{0} & \texttt{25} & \texttt{9.343034507158935e-01} & \texttt{9.343034516458660e-01} \\
		\texttt{1} & \texttt{46} & \texttt{1.226422927658922e+00} & \texttt{1.226422927676387e+00} \\
		\texttt{2} & \texttt{72} & \texttt{1.207656897946988e+00} & \texttt{1.207656897794478e+00} \\
		\texttt{3} & \texttt{104} & \texttt{1.174028026341686e+00} & \texttt{1.174028025751143e+00} \\
		\texttt{4} & \texttt{142} & \texttt{1.125495438561443e+00} & \texttt{1.125495437195381e+00} \\
		\texttt{5} & \texttt{188} & \texttt{1.294921256799873e+00} & \texttt{1.294921252331416e+00}
	\end{tabular}
	\normalsize
	\caption{Starting points for the six continuation branches in the Gaussian well example. $\Re(k)=0$ for all points.}
	\label{tab:gaussstartpoints}
\end{table}

\begin{table}
	\centering
	\footnotesize
	\begin{tabular}{rrrr}
		\hline
		$n$ & $\lambda$ & $\Re(k)$ & $\Im(k)$ \\\hline
		\texttt{0} & \texttt{2.35539E+01} & \texttt{1.06202E-31} & \texttt{1.35574E-05} \\
		\texttt{1} &\texttt{4.28137E+01} & \texttt{-1.05730E-33} & \texttt{-5.66429E-07} \\
		\texttt{2} & \texttt{6.81625E+01} & \texttt{1.75863E-30} & \texttt{4.32438E-07} \\
		\texttt{3} & \texttt{9.96592E+01} & \texttt{-2.41634E-32} & \texttt{1.02708E-04} \\
		\texttt{4} & \texttt{1.37339E+02} & \texttt{1.69583E-25} & \texttt{-1.79228E-05} \\
		\texttt{5} & \texttt{1.81223E+02} & \texttt{3.64142E-26} & \texttt{6.92222E-06}
	\end{tabular}
	\normalsize
	\caption{Branching points of the six branches in the Gaussian well example.}
	\label{tab:gaussbranchpoints}
\end{table}

\begin{figure}
	\centering
	\subfigure[Complex $k$ plane]{\includegraphics[width=3cm]{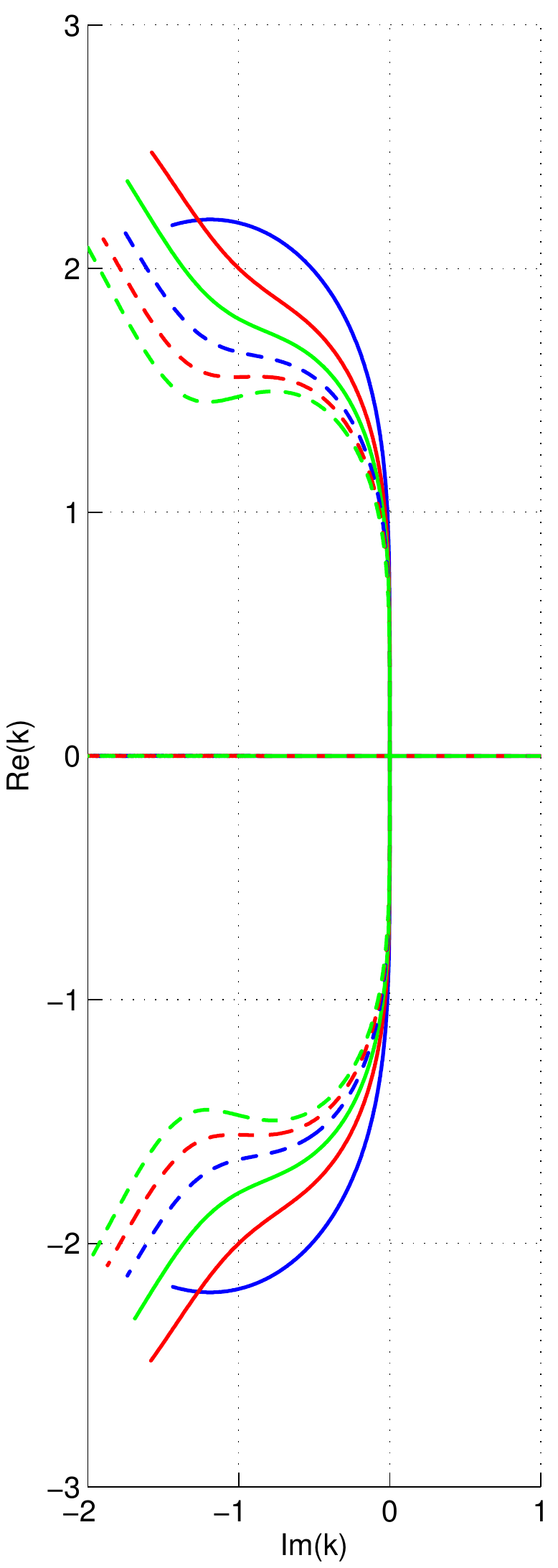}\label{fig:gausscontinuation1}}
	\hspace{1cm}
	\subfigure[$\Im(k)\times\lambda$ plane]{\includegraphics[width=3cm]{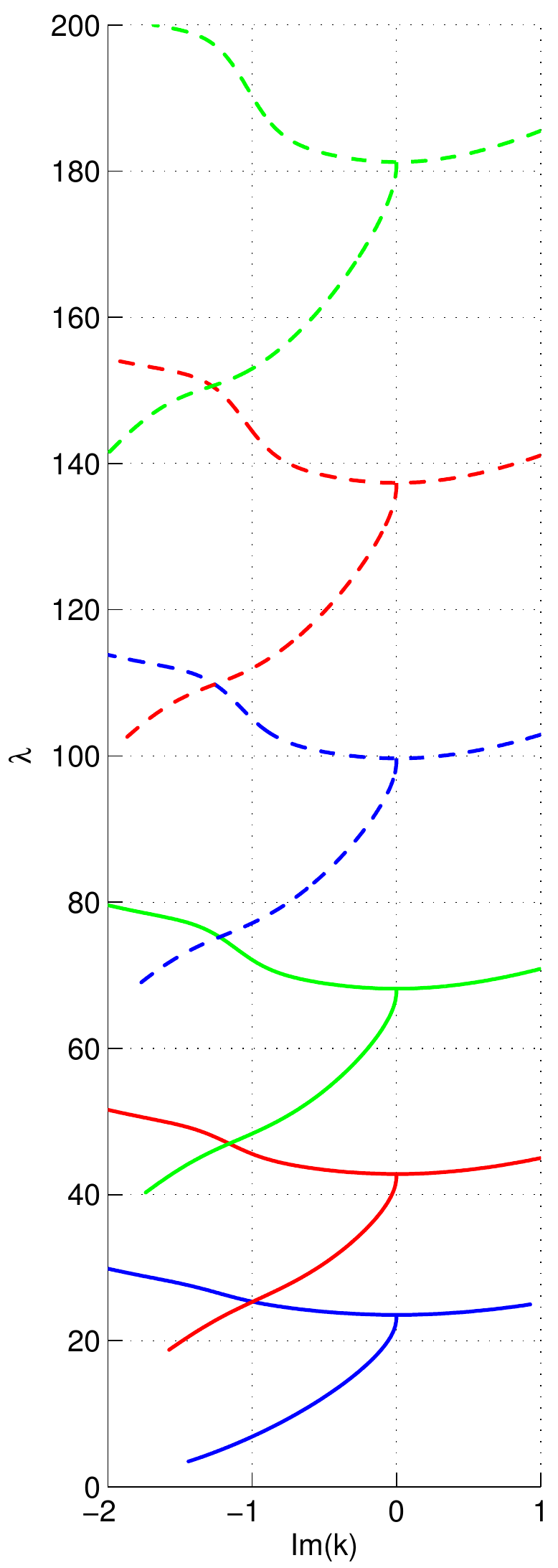}\label{fig:gausscontinuation2}}
	\caption{Projections of trajectories of the $S$-matrix poles representing the first six bound/resonant states for ($l=3$)-waves in a Gauss potential.}
\end{figure}

\begin{figure}
	\centering
	\includegraphics[width=6cm]{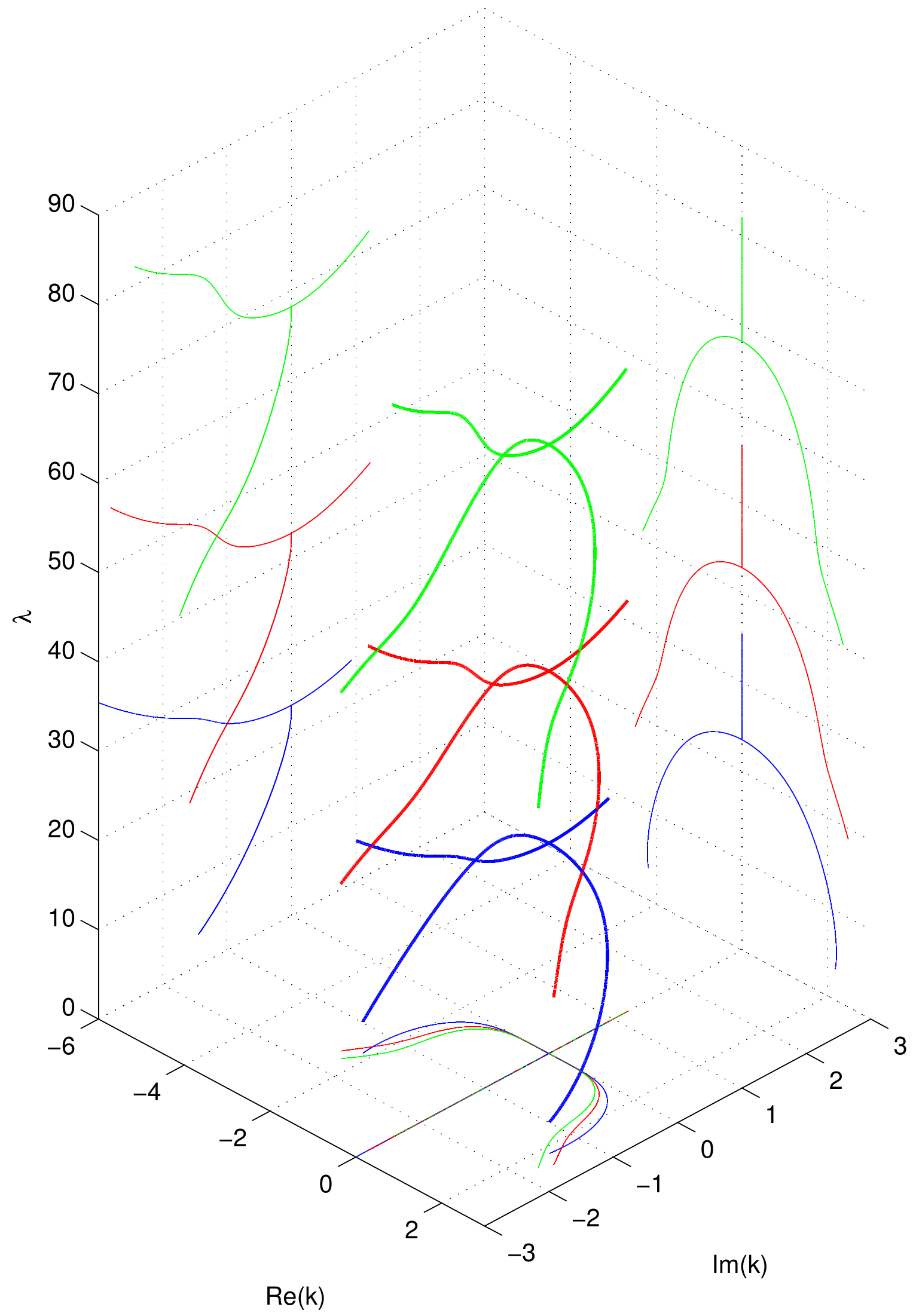}
	\caption{Schematic view of the full trajectories of the $S$-matrix poles representing the first three bound/resonant states for ($l=3$)-waves in a Gauss potential shown in the $k\times\lambda$ space along with projections on the various planes.}
	\label{fig:gausscontinuation3}
\end{figure}

The resulting trajectories of the continuation process are shown in figures \ref{fig:gausscontinuation1}, \ref{fig:gausscontinuation2} and \ref{fig:gausscontinuation3}. The time to compute each trajectory is of the order of several seconds on modern desktop computer hardware, depending on the step size and the number of continuation points.

\subsection{Square potential well}
As a second example we use a $s$-wave ($l=0$) in a square potential well:
\begin{equation}
	V(r,\lambda) =
	\begin{cases}
		-\lambda &\quad r < a \\
		0 &\quad r \geq a
	\end{cases}
\end{equation}
For our purposes we choose $a=1$.

Analytical results for such potentials are well known and were extensively studied in \cite{Nussenzveig1959}. We use them as reference for our numerical studies.

\begin{figure}
	\centering
	\subfigure[Complex $k$ plane]{\includegraphics[width=3cm]{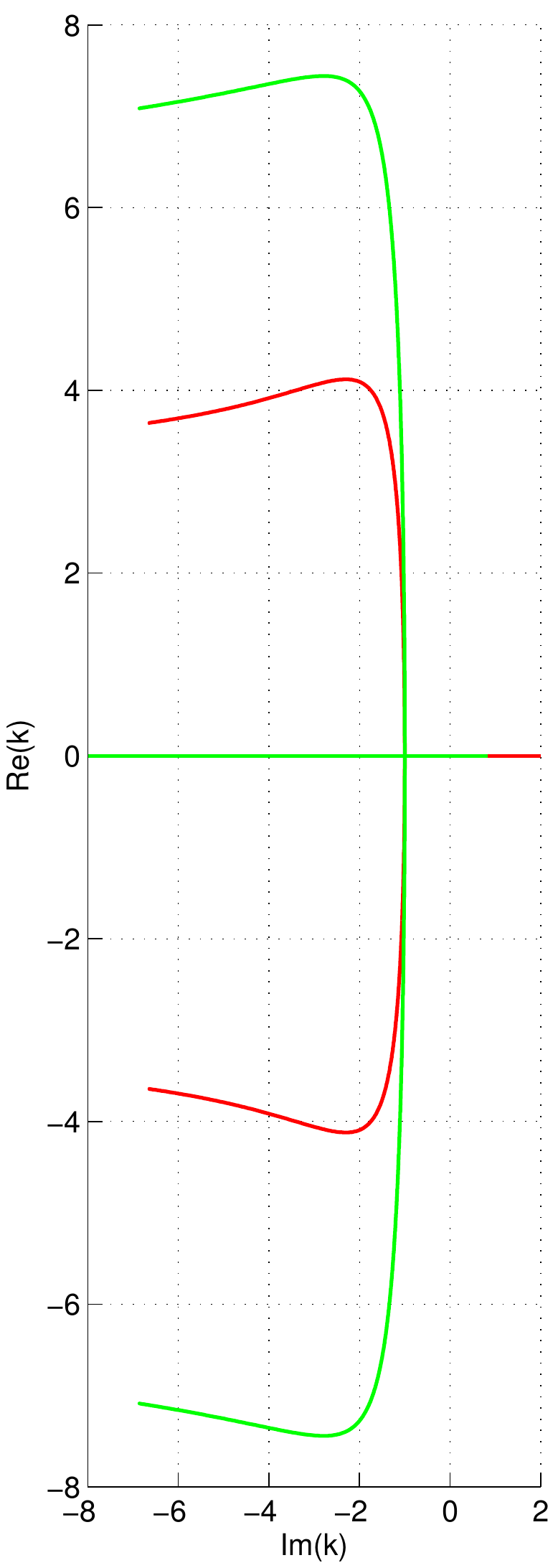}\label{fig:sqwcontinuation1}}
	\hspace{1cm}
	\subfigure[$\Im(k)\times\lambda$ plane]{\includegraphics[width=3cm]{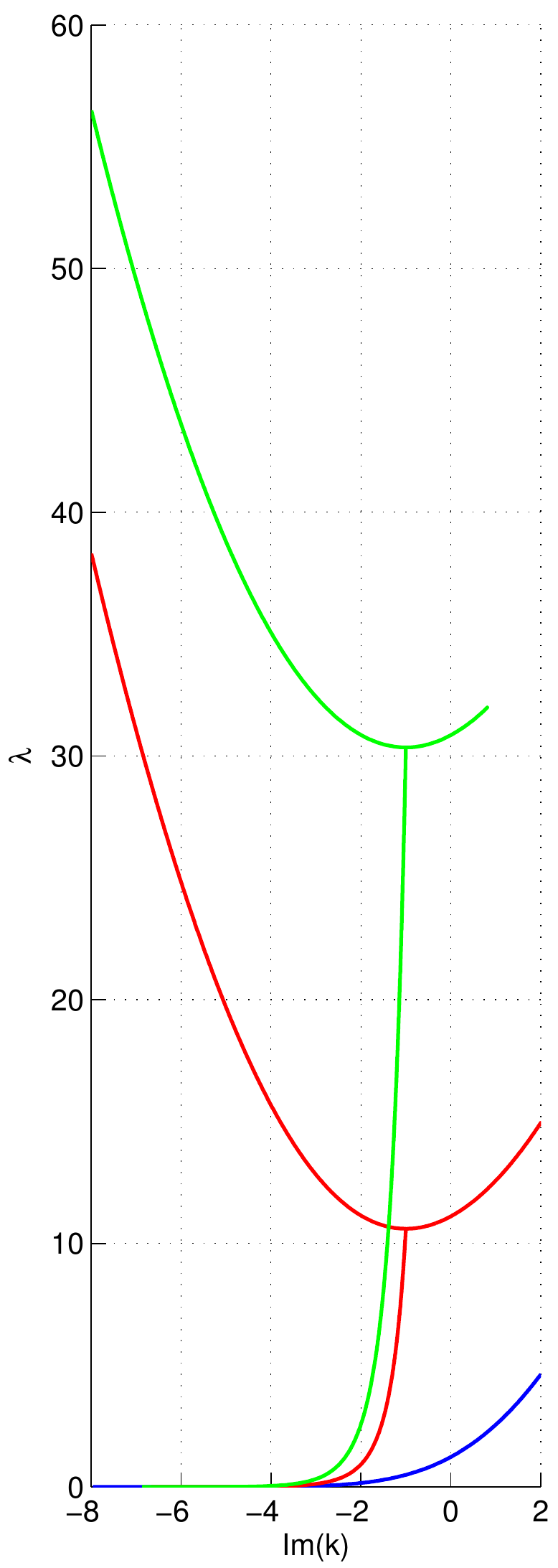}\label{fig:sqwcontinuation2}}
	\caption{Projections of trajectories of the $S$-matrix poles representing the first three bound/resonant states for $s$-waves in a square potential well. Note the theoretically well-known lack of bifurcation in the ground-state branch. The other bifurcations are located at $k=-i$ and not at the origin which is a known result for these types of $s$-wave problems.}
\end{figure}

\begin{figure}
	\centering
	\includegraphics[width=6cm]{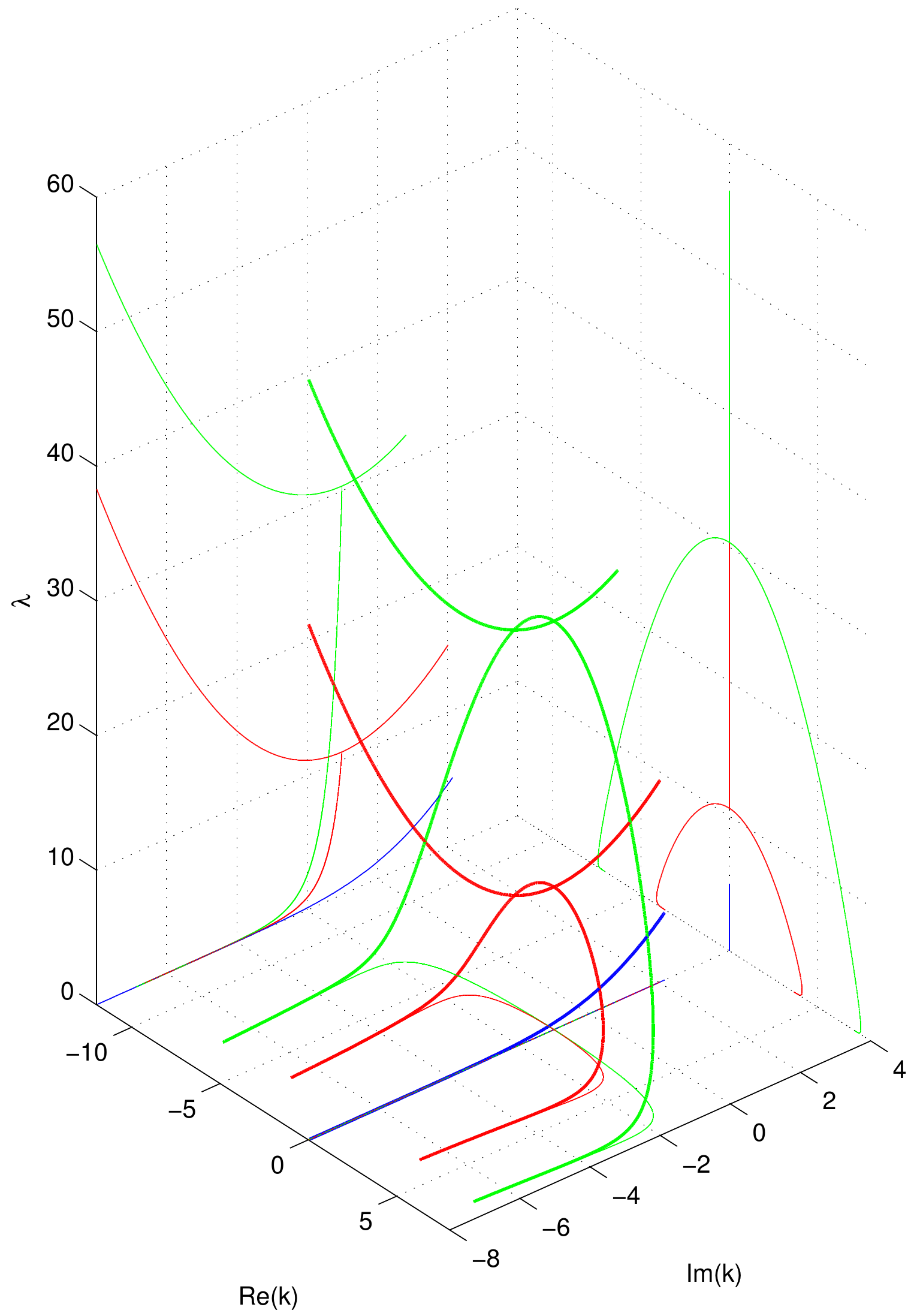}
	\caption{Schematic view of the full trajectories of the $S$-matrix poles for $s$-waves in a square potential well shown in the $k\times\lambda$ space along with projections.}
	\label{fig:sqwcontinuation3}
\end{figure}

The grid used for the renormalized Numerov solver was $r\in[0,\ 1.1]$ with 2048 points and the step size $\Delta s$ was the same as in the previous example. The first three bound states were used for the continuation and the starting points of these three branches are given in table \ref{tab:sqwstartpoints}. As in the Gauss potential case, the continuation was performed from these points in the direction of the origin resulting in branches shown in figures \ref{fig:sqwcontinuation1}, \ref{fig:sqwcontinuation2} and \ref{fig:sqwcontinuation3}. As follows from theoretical considerations in \cite{Nussenzveig1959}, the ground state of this potential does not transform into a resonance. All other states $n>0$ do branch off into resonances at $k=-i$. The real part of $k$ tends to $\pm n\pi$ as $\Im(k) \to -\infty$ and $\lambda\to0$, which again, corresponds to theoretical results.

\begin{table}
	\centering
	\footnotesize
	\begin{tabular}{rrr}
		\hline
		$n$ & $\lambda$ & $\Im(k)$ \\\hline
		\texttt{0} & \texttt{5} & \texttt{2.15040e+00} \\
		\texttt{1} & \texttt{15} & \texttt{2.02173e+00} \\
		\texttt{2} & \texttt{32} & \texttt{8.25262e-01}
	\end{tabular}
	\normalsize
	\caption{Starting points for the three continuation branches in the square potential well example, obtained with the renormalized Numerov method. $\Re(k)=0$ for all points.}
	\label{tab:sqwstartpoints}
\end{table}

%!TEX root = main.tex
\section{Discussion and Conclusions}
\label{sec:conclusions}
Changing system parameters in the potential of a Schr\"odinger equation can turn resonances into bound states or vice versa.  This transition
is usually marked with a double pole of the $S$-matrix. We have interpreted this double pole as a bifurcation point where different states meet.

In this article we have reviewed the key numerical and mathematical
methods that allow us to track these poles and detect the bifurcation
point in an automatic way.  These methods, originally developed by the
dynamical systems community, are based on predictor-corrector methods and are directly applicable to the problem at
hand. What is needed is a numerical routine that solves the radial
Schr\"odinger equation for a given complex $k$ and system parameter
$\lambda$.  The continuation method then calls this routine multiple
times with appropriate arguments $k$ and $\lambda$ and constructs with this
information the complete continuation curve.

Our contribution is the insight to apply the method to a function
proportional to the Jost function  instead of the numerical
$S$-matrix. The latter suffers from nearby zeros and
poles what can lead to diverging and deteriorating behavior.  Another 
insight is to apply the method in  the $k$-plane rather than the
$E$-plane. This gives orthogonal tangent directions in
the bifurcation point while in the $E$-plane these are
aligned and are much harder to treat numerically.

Our approach is quite robust. In addition to the examples provided, we have applied it to other short-range problems including Morse, Yukawa and Lennard-Jones potentials with $l$ ranging from 0 up to 5. In all cases the program worked without any modifications. Also for $s$-wave problems with barriers, where the bifurcation does not happen at the origin, the program was able to detect the bifurcation point, the tangent directions and follow the solution curves that emerge from the bifurcation point. For all these problems the complete curve was found within seconds.

However, we have found it hard to trace resonances when they move far down in 
the complex plane into the region with large negative imaginary
momenta.  In this region, the fundamental solutions $\hat{h}^+_l$ and
$\hat{h}^-_l$ are increasing and decreasing exponentials with a
slight oscillation.  In our shooting method that integrates outwards starting
from $r=0$ the exponentially increasing function will
dominate over the decreasing function. This effect becomes more
difficult to deal with if $R$, the end of the domain, is increased. A
possible solution to this problem might be to use a mismatch function
where two shootings, one from the left and one from the right, are
matched. The shooting from the right would then have $\hat{h}^+_l(kR)$ as boundary condition. It is the question, however, if the
mismatch function is suitable for pseudo-arclength continuation
near the bifurcation point.

The proposed method is, however, independent of this solver and can be
built around any solver of the radial Schr\"odinger equation. In addition to the renormalized Numerov solver, we have tested the method with a finite difference matrix method with
very similar results.

In the future, we will extend the method to problems with unknown
asymptotic solutions that require absorbing boundary conditions such
as ECS~\cite{Rescigno2000}.  This will allow us to track resonances
in multidimensional scattering problems.  Note, that it is then
not possible to solve for the $S$-matrix in the complete complex
$k$-plane.  It is for these large scale problems that our method can
prove to be valuable since other methods, based on interpreting a resonance
as an eigenvalue, then become intractable because they do not scale to a large number of unknowns. Finally, as indicated in section~\ref{sec:qm} the application of our methods to long-range potentials is important as well. This would allow us to tackle problems with a broader range of physical applications and to verify our results with experimental data.

\section*{Acknowledgements}
We gratefully acknowledge support from FWO-Vlaanderen through project number G.0120.08.

%% Reference section:
%\section*{References}

% Use BibTeX:
%\bibliographystyle{elsarticle-num}
% \bibliographystyle{abbrv}
%\bibliography{Refs}

% Or copy-paste main.bbl created by BibTeX here:

\end{document}